# Rapidity distributions of strange particles in Pb-Pb at 158 A GeV/c

Giuseppe E. Bruno on behalf of the NA57 Collaboration[1]

*Dipartimento IA di Fisica dell'Università e del Politecnico di Bari and INFN, Bari, Italy*

**Abstract.** The production at central rapidity of $K^0_S$, $\Lambda$, $\Xi$ and $\Omega$ particles in Pb–Pb collisions at 158 $A$ GeV/*c* has been measured by the NA57 experiment over a centrality range corresponding to the most central 53% of the inelastic Pb–Pb cross section. We present the rapidity distribution of each particle in the central rapidity unit. The distributions are analysed based on hydro-dynamical models of the collisions.

## 1. Introduction

Lattice quantum chromodynamic calculations predict a new state of matter of deconfined quark and gluons (quark gluon plasma, QGP) at an energy density exceeding ~ 1 GeV fm$^{-3}$ [12]. Nuclear matter at high energy density has been extensively studied through ultra-relativistic heavy ion collisions (for recent developments, see [16]).

Within the experimental programme with heavy-ion beams at CERN SPS, NA57 is a dedicated experiment for the study of the production of strange and multi-strange particles in Pb–Pb collisions at mid-rapidity [10].

The measurement of strange particle production provides a fundamental tool to study the dynamics of the reaction. In particular, an enhanced production of strange particles in nucleus–nucleus collisions with respect to proton-induced reactions was suggested long ago as a possible signature of the phase transition from colour confined hadronic matter to a QGP [15]. The enhancement is expected to increase with the strangeness content of the hyperon. These features were first observed by the WA97 experiment [1] and subsequently confirmed and studied in more detail by the NA57 experiment [7]. Other insights into the reaction dynamics have been obtained by NA57 from the study of the $p_T$ distributions of strange particles: the results of the transverse expansion of the collision and the $p_T$ dependence of the nuclear modification factors have been presented, respectively, in [2] and [3].

---

[1] For the author list see http://wa97.web.cern.ch/WA97/NA57authors/index.html

Rapidity distributions provide a tool to study the longitudinal dynamics; for instance differences between protons and anti-protons have been interpreted as a consequence of the nuclear stopping [9]. If hyperons, like protons, keep a 'memory' of the initial baryon density, then the *relative* pattern for the rapidity distribution of hyperons and anti-hyperons should resemble that of protons and anti-protons [5].

Hydrodynamical properties of the expanding matter created in heavy ion reactions have been discussed by Landau [13] and Bjorken [6] in theoretical pictures using different initial conditions. In both scenarios, thermal equilibrium is quickly achieved and the subsequent isentropic expansion is governed by hydrodynamics.

The complete results on the analysis of the rapidity distributions of strange particles can be found in [4]; in this contribution to the Quark Matter 2005 Conference proceedings we discuss mainly the longitudinal dynamics.

**2. Data sample and analysis**

The results presented here are based on the analysis of the full data sample collected in Pb–Pb collisions at 158 $A$ GeV/$c$, consisting of 460 M events. The sample of events corresponds to the most central 53% of the inelastic Pb–Pb cross section. The data sample has been divided into five centrality classes (0, 1, 2, 3 and 4, class 4 being the most central) according to the value of the charged particle multiplicity around central rapidity measured by a silicon microstrip multiplicity detector. The procedure for the measurement of the multiplicity distribution and the determination of the collision centrality for each class is described in [11]. The fractions of the inelastic cross section for the five classes, calculated assuming a total cross section of 7.26 barn, are given in table 1. A detailed description of the particle selection procedure, as well as of the corrections for geometrical acceptance and for detector and reconstruction inefficiencies, can be found in [2,3,4,7]. All the selected $\Xi$ and $\Omega$ hyperon candidates have been individually weighted for acceptance and inefficiency losses; for the much more abundant $K_s^0$ and $\Lambda$ species, the selected particles have been sampled uniformly over the whole data taking period; the sizes of those sub-samples were chosen in order to reach a statistical accuracy better than the limits imposed by the systematic errors. The experimental procedure for the determination of the rapidity distributions is described in [4].

## 3. Strange particle rapidity distributions

The measured rapidity distributions are shown in fig.1 with closed symbols. For all hyperons the rapidity distributions are found to be symmetric with respect to the rapidity of the centre of mass ('mid-rapidity') within the statistical errors as expected for a symmetric collision system. A similar conclusion cannot be drawn for $K_s^0$ since our acceptance coverage does not extend to backward rapidity. The symmetry of the Pb–Pb colliding system allows us to reflect the rapidity distributions around mid-rapidity (open symbols in fig.1). The rapidity distributions of $\Lambda$, $\Xi^-$, $\overline{\Xi}^+$ and $\Omega$ are compatible, within the error bars, with being flat within the NA57 acceptance window.

For the $K_s^0$ and $\overline{\Lambda}$ spectra, on the other hand, we observe a rapidity dependence. The rapidity distributions for these particles are well described by Gaussians centred at mid-rapidity. For both particles, the width of the rapidity distributions is constant within the errors in the five centrality classes (i.e. from 40–53% to 0-4.5%, see table 1).

In all the centrality classes, the rapidity distribution of the Λ hyperon is consistent with being flat over the considered range. In the same rapidity range, the $\overline{\Lambda}$ distribution varies by about 40% (class 4). It is likely that the Λ hyperon rapidity distribution reflects the overall net baryon number distribution. The same behaviour was observed for the *y* distribution of protons in central Pb–Pb collisions at the same energy by the NA49 experiment [5].

## 4. Longitudinal dynamics

The *transverse* dynamics of the collisions have been studied in [2,8] from the analysis of the transverse momentum distributions of strange particles in the framework of the blast-wave model [17]. The rapidity distributions can be used to extract information about the *longitudinal* dynamics. We use an approach outlined in [17] (i.e. the same blast-wave model used for the study of the transverse dynamics) and [14], where, respectively, Bjorken and Landau hydrodynamics [6, 13] are folded with a thermal distribution of the particle velocity in the fluid elements.

In fig.2 the observed rapidity distributions are compared with the expectation for a stationary thermal source and with a longitudinally boost-invariant superposition of multiple isotropic, locally-thermalized sources (i.e. Bjorken hydrodynamics). Each locally thermalized source is modelled by an $m_T$-integrated Maxwell–Boltzmann with the rapidity dependence of the energy, $E = m_T \cosh(\eta)$, explicitly included

$$\frac{dN_{th}}{d\eta} = AT_f^3 \left( \frac{m^2}{T_f^2} + \frac{m}{T_f} \frac{2}{\cosh \eta} + \frac{2}{\cosh^2 \eta} \right) \exp\left( -\frac{m}{T_f} \cosh \eta \right) \qquad (1)$$

where $T_f$ is the freeze-out temperature, $m_T = \sqrt{p_T^2 + m^2}$ and η is the rapidity of the individual fluid element. The distributions are integrated over source element rapidity to extract the maximum longitudinal flow, $\eta_{max}$,

$$\frac{dN}{dy} = \int_{-\eta_{max}}^{\eta_{max}} \frac{dN_{th}}{d\eta}(y-\eta)d\eta \qquad \beta_L = \tanh\eta_{\eta_{max}} \qquad (2)$$

where $\beta_L$ is the maximum longitudinal velocity in units of $c$. The average longitudinal flow velocity is evaluated as $\langle\beta_L\rangle = \tanh(\eta_{max}/2)$. A simultaneous fit of the function defined by equation (2) to the rapidity distributions of the strange particles gives $\langle\beta_L\rangle$=0.42±0.03 with $\chi^2/ndf$=28.2/32. The freeze-out temperature has been fixed to the value $T_f$=144 MeV obtained, for the most central 53% of the inelastic Pb–Pb collisions, from the analysis of the transverse mass spectra of the same group of particles [2]. In the same analysis the average *transverse* flow velocity has been determined to be $\langle\beta_\perp\rangle$=0.38±0.02, i.e. only slightly less than the *longitudinal* velocity determined in this analysis; this indicates substantial stopping of the incoming nuclei as a consequence of the collision.

In principle, also the freeze-out temperature can be fitted from the rapidity distributions along with the longitudinal velocity. However, the sensitivity to the freeze-out temperature is very limited. For instance, changing $T_f$ from 144 to 120 MeV results in only a 2% increase of $\langle\beta_L\rangle$. Within our uncertainties, we do not observe any particle to deviate from the common description given by a collective longitudinal flow superimposed to the thermal motion. A combined fit performed only to the $K_S^0$ and $\overline{\Lambda}$ rapidity distributions yields a smaller value of the flow, i.e. $\langle\beta_L\rangle$=0.36±0.03. It is worth noting that the flattening of the rapidity spectra with increasing particle mass, which is also observed in the data, is due in the model to the collective dynamics: all particles are driven by the flow with the same velocity independently of their masses.

In Landau hydrodynamics, the amount of entropy *(dS)* contained within a (fluid) rapidity d$\eta$ is given by [13]

$$\frac{dS}{d\eta} = \pi R^2 l s_0 \beta c_s \exp[\beta \omega_f] \left[ I_0(q) - \frac{\beta \omega_f}{q} I_1(q) \right] \qquad (3)$$

where $q = \sqrt{\omega_f^2 - c_s^2 \eta^2}$, $\omega_f = \ln(T_f / T_0)$, $c_s$ is the speed of sound in the medium, $T_0$ is the initial temperature, $\eta$ is the rapidity, $R$ is the radius of the nuclei, $2l$ is the initial length, $s_0$ is the initial entropy density, $2\beta = (1 - c_s^2)/c_s^2$ and $I_0$, $I_1$ are Bessel functions. The quantity $\pi R^2 l s_0$ is used to normalize the spectra at mid-rapidity. The particle rapidity distribution is obtained, as for the Bjorken case, as a superposition of the multiplicity density in rapidity space $(dN/d\eta \propto dS/d\eta)$ with a thermal distribution of the fluid elements,

$$\frac{dN}{dy} = \int \frac{dN}{d\eta} \frac{dN_{th}}{d\eta}(y - \eta) d\eta \qquad (4).$$

In the Landau model the width of the rapidity distribution is sensitive to the speed of sound and to the ratio of the freeze-out temperature to the initial temperature. While integrating over $\eta$ in equation (2), the range of $\eta$ is treated as a parameter in the case of Bjorken hydrodynamics; moreover in the Bjorken case (equation (2)) the factor $dN/d\eta$, which appears in equation (4), has been included in the overall normalization factor $A$ since the entropy density $dS/d\eta$ is independent of the rapidity, in accordance with the assumption of boost invariance along the longitudinal direction [6]. In the case of Landau hydrodynamics, the integration limit[2] is fixed by $\eta_{max} = -\eta_{min} = \ln(T_0/T_f)/c_s$ and the multiplicity density in $\eta$ space $(dN/d\eta)$ is written explicitly in the $\eta$ integration (equation (4)). Landau hydrodynamics can also reproduce simultaneously the distributions for all the strange particles considered ($\chi^2$/ndf $\cong$ 28/32), but we are not able to put stringent constraints on both the speed of sound and the ratio $T_f/T_0$. The confidence level contours in the $c_s^2$ versus $T_f/T_0$ parameter space are shown in fig.3. For instance, the hypothesis of a perfect gas (i.e. $c_s^2 = 1/3$) would result (at the $3\sigma$ confidence level) in either $T_f/T_0 \approx 0$ or $T_f/T_0 \approx 0.6$. In fact, two physical regions are constrained at the $3\sigma$ confidence level, the first located at small values of $T_f/T_0$ and the second between 0.5 and 0.8; on the other hand, the region at $c_s^2 > 1/3$ is unphysical. Both physical regions span over the full range $0 < c_s^2 < 1/3$.

---

[2] In [18] a modification has been developed (*Srivastava*) where the integration limit for rapidity is infinite, but this case has not been considered in the present analysis.

## 5. Conclusions

We have measured the d$N$/d$y$ distributions of high purity samples of $K_S^0, \Lambda, \Xi$ and $\Omega$ particles produced at central rapidity in Pb–Pb collisions at 158 $A$ GeV/$c$ over a wide centrality range of collision (i.e. the most central 53% of the Pb–Pb inelastic cross section). In the unit of rapidity around mid-rapidity covered by NA57, we have performed fits to the d$N$/d$y$ distributions of $K_S^0$ and $\overline{\Lambda}$ using a Gaussian parameterization: the resulting widths are compatible with each other and constant as a function of centrality. Contrary to $\overline{\Lambda}$, the $\Lambda$ spectra are flat to good accuracy in the range of rapidity and centrality considered; this would indicate that the $\Lambda$ hyperon conserves 'memory' of the initial baryon density.

The rapidity distributions of the $\Omega$ particle are found to be flat within the errors in one unit of rapidity for central (0–11%) and peripheral (23–53%) collisions.

Boost-invariant Bjorken hydrodynamics can describe simultaneously the rapidity spectra of all the strange particles under study with $\chi^2$/ndf $\approx 1$, yielding an average longitudinal flow velocity $\langle \beta_L \rangle = 0.42 \pm 0.03$, slightly larger than the measured transverse flow. An almost *isotropic* collective expansion of the system suggests large nuclear stopping.

A fairly good description is also provided by Landau hydrodynamics, which allows us to put constraints in the parameter space of the speed of sound in the medium and the ratio of the freeze-out temperature to the initial temperature.

**List of figures.**

1. Rapidity distributions of strange particles in the most central 53% of Pb–Pb interactions at 158 *A* GeV/*c*. Closed symbols are measured data and open symbols are measured points reflected around mid-rapidity. The $\overline{\Lambda}$ and $\overline{\Xi}^+$ results have been scaled by factors 4 and 2, respectively, for display purposes. The superimposed boxes show the yields measured in one unit of rapidity (as published in [2]) with the dashed and full lines indicating the statistical and systematic errors, respectively.

2. Rapidity distributions of strange particles for the centrality range corresponding to the most central 53% of the inelastic Pb–Pb cross section as compared to the thermal model calculation of equation (1) (dotted lines, in red) and a thermal model with longitudinal flow (full lines, in black).

3. The square of the speed of sound in the medium (in unit of $c^2$) versus the ratio of the freeze-out temperature to the initial temperature. The 1σ (full curves) and the 3σ (dashed curves) confidence contours are shown. The dotted line at $c_s^2 = 1/3$ shows the ideal gas limit.

**Table 1.** Centrality ranges for the five classes.

| Class | 0 | 1 | 2 | 3 | 4 |
|---|---|---|---|---|---|
| $\sigma/\sigma_{inel}$ (%) | 40–53 | 23–40 | 11–23 | 4.5–11 | 0–4.5 |

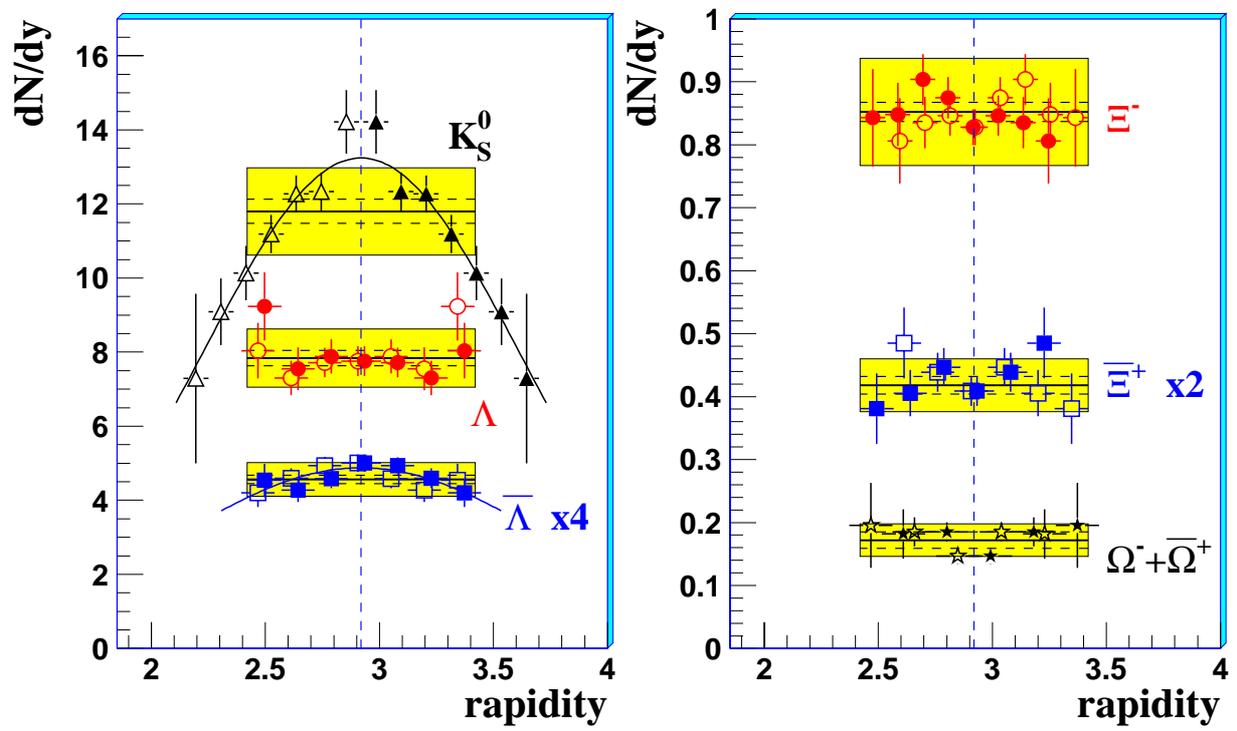

Figure 1

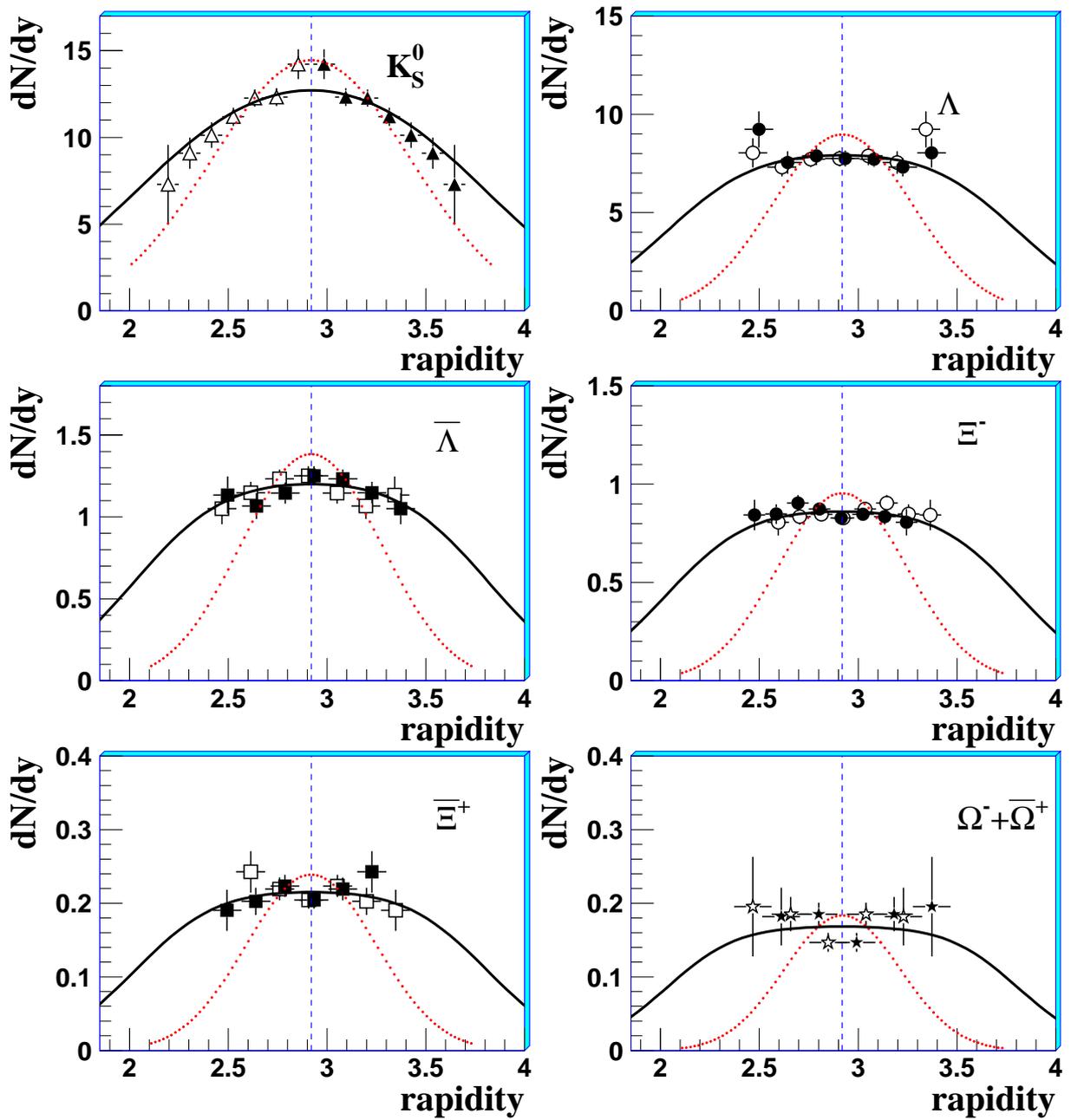

Figure 2

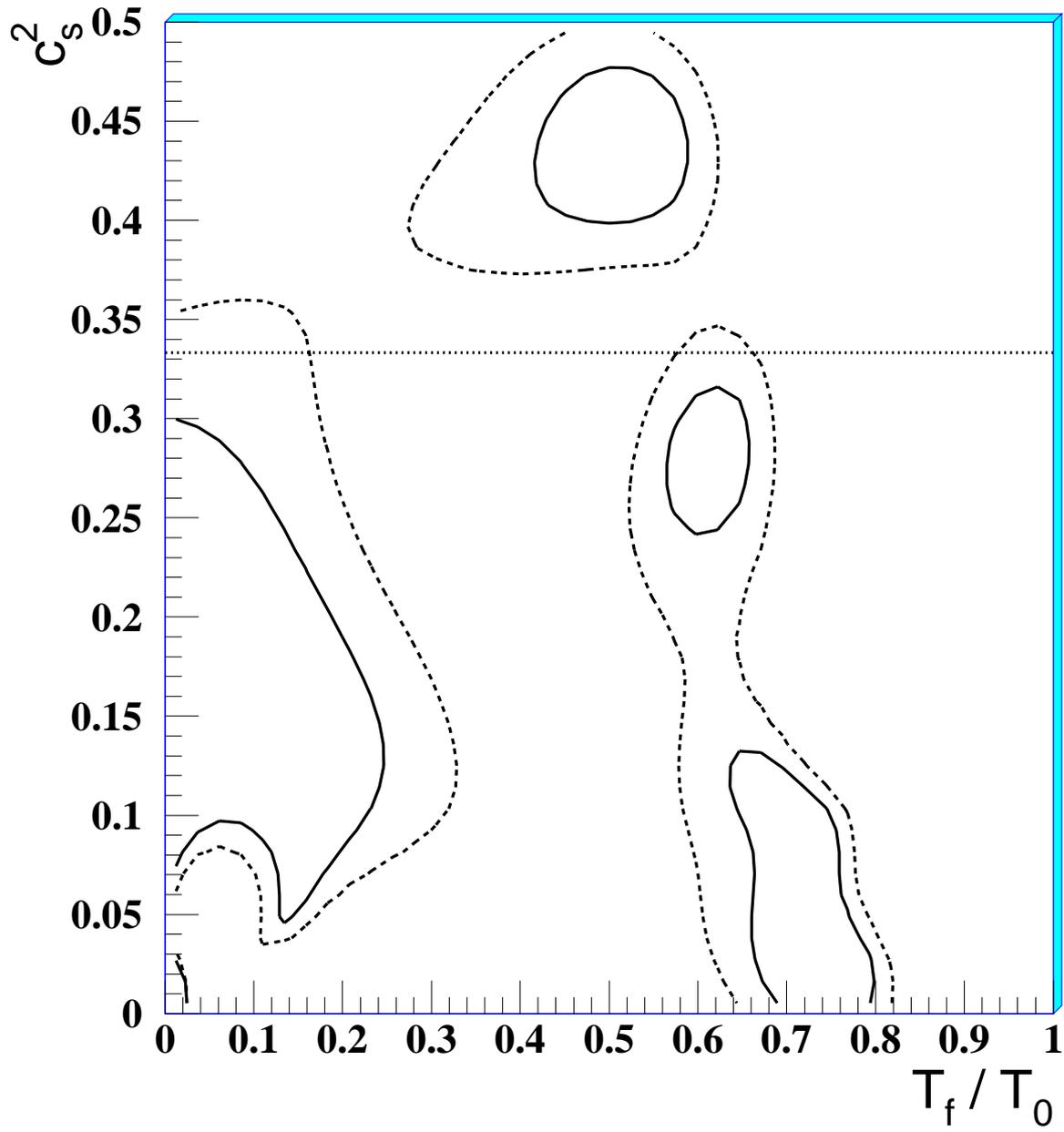

Figure 3